\documentclass[aps,prb,onecolumn,showpacs,showkeys]{revtex4}

\usepackage{epsfig}
\usepackage{amsmath}
\usepackage{amsfonts}
\usepackage{amssymb}
\usepackage{slashed}
\usepackage{graphicx}
\usepackage{subfigure}
\usepackage[latin1]{inputenc}
\newcommand{\be}{\begin{equation}}
\newcommand{\ee}{\end{equation}}

\begin{document}

\title{A note on the analogy between superfluids and cosmology}
\author{A. Naddeo\footnote{e-mail: \textit{naddeo@sa.infn.it}}$^{(1)}$ and G. Scelza\footnote{e-mail: \textit{Giovanni.Scelza@unige.ch}}$^{(2)}$}
\affiliation{(1) CNISM, Unit\`{a} di Ricerca di Salerno and
Dipartimento di Fisica ``E. R. Caianiello'', Universit\`{a} degli
Studi di Salerno, Via Salvador Allende, $84081$ Baronissi
(SA), Italy \\
(2) Universit\'{e} de Gen\`{e}ve, Departement de Physique
Theorique, $24$, quai E. Ansermet, CH-$1211$ Gen\`{e}ve,
Switzerland}

\date{\today}

\begin{abstract}
A new analogy between superfluid systems and cosmology is here
presented, which relies strongly on the following ingredient: the
back-reaction of the vacuum to the quanta of sound waves. We show
how the presence of thermal phonons, the excitations above the
quantum vacuum for $T> 0$, enable us to deduce an hydrodynamical
equation formally similar to the one obtained for a perfect fluid
in a Universe obeying the Friedmann-Robertson-Walker metric.
\end{abstract}
 \maketitle
\section{Introduction}
Different condensed matter systems, such as acoustics in flowing
fluids, light in moving dielectrics or quasiparticles in moving
superfluids, can be shown to reproduce some aspects of General
Relativity (GR) and cosmology \cite{volov,liberati}. They can be
conceived as laboratory toy models in order to make experimentally
accessible some features of quantum field theory on curved-space.
The starting point of such an exciting and fruitful research area
can be found in the celebrated \emph{acoustic black hole} by Unruh
\cite{Unruh}. Here the analogy between the motion of sound waves
in a convergent fluid flow and massless spin-zero particles
exposed to a black hole was first outlined. Since then, the search
for an emergent space-time has been extended to various media,
such as electromagnetic waveguides \cite{Unruh1}, superfluid
helium \cite{volov} and Bose-Einstein condensates
\cite{liberati1}. Emergent space-time and gravity effects in
superfluids are of particular interest. Indeed the extremely low
temperatures experimentally accessible allow in principle the
detection of tiny quantum effects, such as Hawking radiation,
particle production and quantum back-reaction
\cite{volo,fischer1}. Bose-Einstein condensates made of cold atoms
in optical lattices are very promising because of the high degree
of experimental control \cite{optical,optical1}. Indeed such
systems have been proposed to mimic an expanding Friedmann,
Robertson, Walker (FRW) universe \cite{wei}, where the behavior of
quantum modes has been reproduced by manipulating the speed of
sound through external fields via Feshbach resonance techniques
\cite{Bar}.

On the other hand the simulation of some gravity effects in
condensed matter systems leads to new insights into the deep
connection between quantum hydrodynamics and quantum gravity
\cite{volo}. In this context one of the main open problems is the
correct treatment of quantum fluctuations, always present on the
top of the classical background which describes the macroscopic
behavior of superfluids. This quantum back-reaction is also
related to fundamental issues such as the Big Bang singularity and
the cosmological constant.

In this letter we take a step forward in the introduction of a new
analogy between superfluid systems and gravity, which relies
strongly on the analysis of the back-reaction of the vacuum to the
quanta of sound waves. In particular, we show how the presence of
thermal phonons, the excitations above the quantum vacuum for $T>
0$, allows us to justify an hydrodynamical equation formally
similar to the one obtained for a perfect fluid in a Universe
obeying the FRW metric \cite{wei}. The letter is organized as
follows. In Section 2 we briefly recall the derivation of the
Friedmann fluid equation within the FRW cosmological model. In
Section 3 we show that the non-zero temperature and the
back-reaction of quanta of sound waves onto the quantum vacuum
allow us to derive a cosmological-like equation. Finally, in
Section 4 some comments and outlooks of this work are given.
\section{The cosmological fluid equation}
In this Section we briefly review how to derive the Friedmann
fluid equation within the FRW cosmological model \cite{wei}.

As a starting point our Universe is assumed homogeneous and
isotropic and this hypothesis is perfectly compatible with
observations on large scales of length $\sim 4000 Mps$. Such an
Universe is described by means of the FRW metric tensor:
$$[g_{\mu\nu}]=\begin{pmatrix}1&0&0&0\\
0&-\frac{a^2(t)}{1-kr^2}&0&0\\
0&0&-a^2(t)r^2&0\\
0&0&0&-a^2(t)r^2\sin^2\vartheta \end{pmatrix},$$ where
$k=+1,\,0,\,-1$ is the sign of the curvature and $a(t)$ is the
expansion factor (case of non-stationary Universe), which is a
function of the time alone in order to have a Universe homogeneous and isotropic.\\
The dynamical link between the matter content of the Universe and
the metric tensor is codified in the Einstein's equations:
\begin{equation}\label{eqn:GR}
  G_{\mu\nu}\equiv R_{\mu\nu}-\frac{1}{2}g_{\mu\nu}R,
\end{equation}
with $G_{\mu\nu}=\frac{8\pi G}{c^4}T_{\mu\nu}$; here $c$ is the
light speed, $G$ is the gravitational constant, $T_{\mu\nu}$ is
the energy-momentum tensor, $R_{\mu\nu}$ is the Ricci curvature
tensor, $R$ is the scalar curvature and $g_{\mu\nu}$ is the metric
tensor. Einstein's equations tell us that \emph{the presence of
matter bends space-time}. In the particular case in which there is
absence of matter we have $T_{\mu\nu}=0$. For a perfect fluid,
that is a fluid which has no viscosity or heat flow, the
energy-momentum tensor can be written as:
$$T_{\mu\nu}=\biggl(\frac{p}{c^2}+\rho\biggr)u_\mu
u_\nu-pg_{\mu\nu},$$ where $p$ is the pressure, $\rho$ is the
density and $u$ the fluid velocity. In a co-moving system, that is
a system at rest with respect to the cosmic fluid, the fluid
velocity is $u=(c,0,0,0)$ and then:
\begin{equation}\label{eqn:enimp}
  T_{00}=\rho c^2,\qquad
T_{11}=\frac{p\,a^2(t)}{1-k\,r^2},\qquad T_{22}=p\,r^2a^2(t),
\qquad T_{33}=p\,r^2a^2(t)\sin^2\vartheta.
\end{equation}
Here $T_{00}$ is the energy density while $T_{11}$, $T_{22}$ and
$T_{33}$ give rise to the pressure. From \ref{eqn:enimp} it
follows that:
$$[T^\mu_{\phantom{\mu}\nu}]=\begin{pmatrix}c^2\rho&0&0&0\\
0&-p&0&0\\
0&0&-p&0\\
0&0&0&-p \end{pmatrix}.$$ Let us now obtain the Friedmann fluid
equation. In order to pursue this task, let us remember that the
Einstein's field equation must satisfy the \emph{Bianchi identity}
$$D_{\nu}G^{\mu\nu}\equiv D_{\nu}T^{\mu\nu}\equiv T^{\mu\nu}_{\phantom{\mu\nu};\nu}=0,$$
where $D_{\nu}$ is the \textit{covariant derivative} and the
rising of the two indices in $G_{\mu\nu}$ and in $T_{\mu\nu}$ is
obtained by means of two applications of the metric tensor
$g^{\mu\nu}=g_{\mu\nu}^{-1}.$ So, we get:
\begin{equation}\label{eqn:dercovt}
  T^{\mu\nu}_{\phantom{\mu\nu};\nu}=\frac{\partial
T^{\mu\nu}}{\partial
x^{\nu}}+\Gamma^{\mu}_{\phantom{\mu}\nu\sigma}T^{\sigma\nu}+
\Gamma^{\nu}_{\phantom{\mu}\nu\sigma}T^{\mu\sigma}=0.
\end{equation}
The \emph{connection coefficients} or \emph{Christoffel symbols},
which are not tensors, $\Gamma^{\mu}_{\phantom{\mu}\nu\sigma}$ are
given by
$$\Gamma^{\mu}_{\phantom{\mu}\nu\sigma}=
\frac{1}{2}g^{\mu\lambda}\biggl(\frac{\partial
g_{\lambda\nu}}{\partial x^{\sigma}}+\frac{\partial
g_{\lambda\sigma}}{\partial x^{\nu}}-\frac{\partial
g_{\nu\sigma}}{\partial x^{\lambda}}\biggr).$$ In order to
understand their physical meaning, let us analyze the
\emph{parallel transport} of an arbitrary contravariant vector
$A^{\alpha}$ in a curved space-time, which describes the Universe
in General Relativity. In particular if its value at a point
$x^\alpha$ is $A^{\alpha}$, then at the neighboring point
$x^{\alpha}+dx^{\alpha}$ it is equal to $A^{\alpha}+dA^{\alpha}$.
Now let the vector $A^{\alpha}$ perform an infinitesimal parallel
displacement to the point $x^{\alpha}+dx^{\alpha}.$ As a result of
such an operation, the change in the vector $A^{\alpha}$ is
denoted by $\delta A^{\alpha}$. Then the difference $DA^{\alpha}$
between the two vectors which are now located at the same point
is:
\begin{equation}\label{eqn:defdercov}
  DA^{\alpha}=dA^{\alpha}-\delta A^{\alpha}.
\end{equation}
It is possible to show that\cite{lan} $\delta
A^{\alpha}=-\Gamma^{\alpha}_{\phantom{\mu}\mu\nu}A^{\mu}dx^{\nu}.$
For a galilean coordinate system the following relation holds:
$\Gamma^{\alpha}_{\phantom{\mu}\mu\nu}=0.$ We explicitly note that
the equation (\ref{eqn:defdercov}) gives us the definition of the
covariant derivative of a contravariant vector once we divide both
sides by $dx^{\nu}$. Furthermore the generalization to a two
indices tensor is provided by Eq. (\ref{eqn:dercovt}).\\
Starting from Eq. (\ref{eqn:dercovt}), after some manipulations we
finally obtain the fluid equation:
\begin{equation}\label{eqn:fluido}
  \dot{\rho}+3\frac{\dot{a}}{a}\biggl(\rho+\frac{p}{c^2}\biggr)=0 \qquad
   \Rightarrow \qquad \frac{d\,\rho}{d\,a}=-\frac{3}{a}\biggl(\rho+\frac{p}{c^2}\biggr).
\end{equation}
Now we wonder if and under which hypothesis it is possible to
obtain a similar equation in a quantum hydrodynamics context. We
address this point in the following Section.
\section{One more similitude}
In this Section we perform a step forward in the introduction of a
new analogy between superfluids and cosmology by working in a
quantum hydrodynamics context. Within such a context, we need to
carefully include in our analysis the effect of the back-reaction
of the vacuum to the quanta of sound waves at non-zero temperature
which leads to a depletion of the mass density $\rho $. In
particular, we are going to show how the presence of thermal
phonons, the excitations above the quantum vacuum for $T>0$,
allows us to justify an hydrodynamical equation formally similar
to the one obtained for a perfect fluid in a Universe obeying the
FRW metric \cite{wei}.

In order to clarify the physical picture behind our derivation let
us briefly recall the main concepts on which Landau's formulation
of quantum hydrodynamics strongly relies \cite{landau}. The
starting point is the quantum Hamiltonian
\begin{equation}
H\left( \widehat{\rho },\widehat{\mathbf{v}}\right) =\int d^{3}x\left( \frac{%
1}{2}\widehat{\mathbf{v}}\widehat{\rho
}\widehat{\mathbf{v}}+\epsilon \left( \widehat{\rho }\right) -\mu
\widehat{\rho }\right) ,  \label{landau1}
\end{equation}
where $\epsilon$ is the energy density of the liquid, $\mu$ is the
chemical potential and the quantum operators $
\widehat{\mathbf{v}}$ (velocity field operator) and $\widehat{\rho
}$ (mass density operator) satisfy the following commutation
rules:
\begin{equation}
\left[ \widehat{\rho }\left( \mathbf{r}_{1}\right) ,\widehat{\rho
}\left( \mathbf{r}_{2}\right) \right] =0,  \label{landau2}
\end{equation}
\begin{equation}
\left[ \widehat{\mathbf{v}}\left( \mathbf{r}_{1}\right) ,\widehat{\rho }%
\left( \mathbf{r}_{2}\right) \right] =\frac{\hbar }{i}\nabla
\delta \left( \mathbf{r}_{1}-\mathbf{r}_{2}\right) ,
\label{landau3}
\end{equation}
\begin{equation}
\left[ \widehat{v}_{i}\left( \mathbf{r}_{1}\right)
,\widehat{v}_{j}\left( \mathbf{r}_{2}\right) \right] =\frac{\hbar
}{i\widehat{\rho }}\varepsilon _{ijk}\left( \nabla \times
\widehat{\mathbf{v}}\right) _{k}\delta \left(
\mathbf{r}_{1}-\mathbf{r}_{2}\right) .  \label{landau4}
\end{equation}
Now let us observe that quantum hydrodynamics is characterized by
the following dimensional quantities: the equilibrium values of
$\rho$ and $c_{s}$ where $c_{s}$ is the speed of sound
(\emph{i.e.} phonons), and the Planck constant $\hbar $. By means
of these three quantities it is possible to build up the
characteristic scales for the energy $E_{QH}$, the mass $M_{QH}$,
the length $a_{QH}$, the frequency $\omega _{QH}$ and the energy
density $\epsilon _{QH}$ within the quantum hydrodynamical
context:
\begin{equation}
\begin{array}{ccccc}
E_{QH}^{4}=\frac{\hbar ^{3}\rho }{c_{s}}, \qquad
M_{QH}^{4}=\frac{\hbar ^{3}\rho }{c_{s}^{3}}, \qquad
a_{QH}^{4}=\frac{\hbar }{\rho c_{s}}, \qquad \omega _{QH}=\left(
\frac{c_{s}^{5}\rho }{\hbar }\right) ^{1/4}, \qquad \epsilon
_{QH}\sim \epsilon \left( \rho \right) \sim \rho c_{s}^{2}
\end{array}
.  \label{landau5}
\end{equation}
As Landau pointed out \cite{landau}, the low energy modes present
in quantum hydrodynamics are only phonons while the rotational
modes (\emph{i.e.} vortices) are separated by a gap. Such a gap is
given by the characteristic energy scale $E_{QH}$ above defined.
Within the linear regime (and in the absence of rotational degrees
of freedom) sound waves are quantized and the phonons obtained
have a linear spectrum $E_{k}=\hbar\,c_{s}k$. In the low energy
limit the superfluid quantum vacuum behaves as a classical liquid
and the quantum fluctuations of the phonon field on the top of
this classical background, albeit small, have some influence on
its dynamics. Indeed they give rise to the depletion of the mass
density $\rho$ of the vacuum which turns out to be an universal
phenomenon \cite {volo}. This is the well known quantum
back-reaction of the vacuum to the phonons which now we quantify.

Indeed the physical picture is the following. At temperature $T>0$
the liquid is made of vacuum with density $\rho$ and phonon
excitations. So it is possible to show how the presence of thermal
phonons modifies the mass density of the quantum vacuum and to
quantify such a density variation. Let us consider very low
temperatures $T\ll E_{QH}$ , so that only low-frequency phonons
with linear spectrum $\omega =k\,c_{s}$ contribute to the thermal
energy. Let us also suppose a fixed external pressure. These
hypotheses imply that the free energy is the sum of two
contributions: the energy of the quantum vacuum and the free
energy of the ``matter'', the phonons. In this case it is possible
to show \cite{volo} that
\begin{equation*}
F(T,\rho )=F_{vac}+F_{mat}=F_{vac}-P_{mat}=\varepsilon (\rho )-\mu \rho -%
\frac{1}{3}\varepsilon _{mat}(\rho ),
\end{equation*}
where $\varepsilon -\mu \rho =\varepsilon _{vac}$ and $\varepsilon
_{mat}$ are, in this order, the \emph{energy density of the
quantum vacuum} and the \emph{energy density of the gas of thermal
phonons} (radiation energy). We indicate with $\rho _{0}$ the
equilibrium density and $\mu _{0}$ be the chemical potential at
$T=0$; then, since $\varepsilon _{mat}$ is considered as a
perturbation, we can expand the free energy $F$ in terms of
$\delta \rho =\rho -\rho _{0}$ and $\delta \mu =\mu -\mu _{0}$
(where $\rho =\rho (T\neq 0)$ and $\mu =\mu (T\neq 0)$). Let us
remember that the chemical potential $\mu $ must be changed in
order to keep a fixed external pressure; furthermore the total
change of the pressure of the liquid, which is given by the vacuum
pressure of the liquid and the radiation pressure of phonons must
be equal to zero.

By taking into account the above considerations and by minimizing over $%
\delta \rho $ the expression of $F$ just obtained we arrive to the
following hydrodynamical equation\cite{volo}
\begin{equation}
\frac{\delta \rho }{\rho }=-\frac{\varepsilon _{mat}}{\rho c_{s}^{2}}\biggl(%
\frac{1}{3}+\frac{\rho }{c_{s}}\frac{\partial c_{s}}{\partial \rho
}\biggr), \label{eqn:fir}
\end{equation}
which is equivalent to
\begin{equation}
\frac{dc_{s}}{d\rho }=-\frac{c_{s}}{3\rho
}\biggl(1+3c_{s}^{2}\frac{\delta \rho }{\varepsilon
_{mat}}\biggr).  \label{eqn:fluid_1}
\end{equation}
Now we try to rewrite Eq. (\ref{eqn:fluid_1}) in a way as similar
as possible to Eq. (\ref{eqn:fluido}). By inverting Eq.
(\ref{eqn:fluid_1}), we can write:
\begin{equation}\label{eqn:presv}
  \frac{d\rho }{dc_{s}}=-\frac{3\rho
}{c_{s}}\biggl(1+3c_{s}^{2}\frac{\delta \rho }{\varepsilon
_{mat}}\biggr)^{-1}.
\end{equation}
Furthermore let us suppose $\frac{\delta \rho }{\varepsilon
_{mat}}<<1$, \emph{i.e.} we suppose, remembering the definition of
$\delta\rho$, that the temperatures we are dealing with will give
rise to a very small variation of the density. \emph{As a
consequence the temperatures are presumably very close to $T=0$}.
In this way, starting from Eq. (\ref{eqn:presv}) and by means of a
series expansion, we get:
\begin{equation}
\frac{d\rho }{dc_{s}}\simeq -\frac{3\rho }{c_{s}}\biggl(1-3c_{s}^{2}\frac{%
\delta \rho }{\varepsilon
_{mat}}\biggr)=-\frac{3}{c_{s}}\biggl(\rho
-3c_{s}^{2}\frac{\rho \,\delta \rho }{\varepsilon _{mat}}\biggr)=-\frac{3}{%
c_{s}}\biggl(\rho -c_{s}^{2}\frac{\rho \,\delta \rho
}{P_{mat}}\biggr)\doteq -\frac{3}{c_{s}}\biggl(\rho +\zeta (\rho
)\biggr),  \label{eqn:fluid_2}
\end{equation}
with a suitable definition of variables. In this way Equations
(\ref {eqn:fluido}) and (\ref{eqn:fluid_2}) look very
similar.\newline Let us notice that the correspondence between
these two equations is
\begin{equation}
c_{s}\leftrightsquigarrow a  \label{eqn:csan}
\end{equation}
and no longer
\begin{equation}
c_{s}\leftrightsquigarrow \frac{1}{a^{2}},  \label{eqn:csao}
\end{equation}
as obtained by Barcel\'{o} et al. \cite{Bar} upon comparing the
spatially flat FRW metric \cite{wei}
($ds_{FRW}^{2}=-c^{2}dt^{2}+\left[ a\left( t\right) \right]
^{2}d\overrightarrow{x}^{2}$) with the Unruh metric \cite {Unruh}
whose components are
\begin{equation*}
g_{00}=-\frac{\rho
(\mathbf{r},t)}{c_{s}}(c_{s}^{2}-\mathbf{v}^{2}),\quad
g_{ij}=\frac{\rho (\mathbf{r},t)}{c_{s}}\delta _{ij},\quad
g_{0i}=-g_{ij}v^{j},
\end{equation*}
where $\mathbf{v}=\overrightarrow{v}\left(
t,\overrightarrow{x}\right) $ is the physical velocity of the
medium (or superfluid) with respect to the laboratory. In the case
of $\mathbf{v}=0,$ then for an inner observer, it is
$g_{\mu \nu }=\mbox{diag}(-1,c_{s}^{-2},c_{s}^{-2},c_{s}^{-2}),$ the \emph{%
Minkowskian acoustic metric}\cite{volov}. We note explicitly that
the result (\ref{eqn:fir}), from which we deduce
(\ref{eqn:fluid_2}), can be obtained from the analysis of
classical hydrodynamic equation made by Stone \cite {stone} for
which the Unruh metric holds. Nevertheless, the conclusion (\ref
{eqn:fluid_2}) could be derived in an alternative way starting
from the effective metric for the superfluid.

We will deal with such a derivation in a forthcoming publication
\cite{noi}.
\section{Conclusions and perspectives}
In conclusion, the arguments given by Volovik \cite{volov} in
order to justify the deep connections and the analogy between
superfluid dynamics and cosmology are here enriched with a new
ingredient: the non-zero temperature and the back-reaction of
quanta of sound waves onto the quantum vacuum. We have been shown
that, in this new physical situation, a cosmological-like equation
can be derived in a natural way. Let us now remember that
measurements of the cosmic microwave background radiation give us
the following value for the temperature of our Universe:
$T_{U}=2.735\,K$, a value very close to $0\,K$. It is interesting
to observe how a quantum fluid at low temperature and with a very
little variation of the density with respect to the temperature
gives rise to an equation formally similar to the fluid equation
of an Universe with low temperature and density
$\rho=\frac{3H_0^2\Omega_0}{8\,\pi\,G}\simeq
9.7\times10^{-27}Kg/m^3$, where $G$ is the gravitational constant,
$H_0$ is the Hubble Constant (here supposed equal to its best fit
value $=72\,Km\times\,s^{-1}\times\,Mpc^{-1}$), and $\Omega_0$ is
the today total density parameter, which for a flat Universe (i.
e. our Universe) it is equal to $1$.

Finally, concerning Equation (\ref{eqn:fluid_2}), we argue that it
is possible to deduce it in an alternative way from the effective
Unruh metric for the superfluid. That will be the subject of a
forthcoming publication \cite{noi}.
\section{Acknowledgements}
We wish to thank M. Grimaldi for valuable discussions and
suggestions.


\begin{thebibliography}{99}
\bibitem{volov} G. Volovik, \textit{The Universe in a Helium Droplet}, Clarendon
Press, Oxford, 2003.
\bibitem{liberati} C. Barcel\'{o}, S. Liberati, M. Visser,
\textit{Analogue Gravity}, Living Rev. Rel. \textbf{8}, 12 (2005).
\bibitem{Unruh} W. G. Unruh, \textit{Experimental black-hole
evaporation?}, Phys. Rev. Lett. \textbf{46}, 1351 (1981).
\bibitem{Unruh1} W. G. Unruh, \textit{Hawking radiation in an electro-magnetic wave-guide?}, Phys. Rev. Lett. \textbf{95}, 1 (2005).
\bibitem{liberati1} C. Barcel\'{o}, S. Liberati, M. Visser,
\textit{Analog gravity from Bose-Einstein condensates}, Class.
Quant. Grav. \textbf{18}, 1137 (2001).
\bibitem{volo} G. Volovik, \textit{From quantum hydrodynamics to quantum
gravity}, in Proceedings of the Eleventh Marcel Grossmann Meeting
on General Relativity, edited by H. Kleinert, R. T. Jantzen and R.
Ruffini, World Scientific, Singapore, 2007; gr-qc/0612134.
\bibitem{fischer1} U. R. Fischer,
\textit{Dynamical aspects of analogue gravity: The backreaction of
quantum fluctuations in dilute Bose-Einstein condensates}, Lect.
Notes Phys. \textbf{718}, 93 (2007).
\bibitem{optical} C. J. Pethick, H. Smith, \textit{Bose-Einstein condensates in dilute gases},
Cambridge University Press, Cambridge, 2008.
\bibitem{optical1} M. H. Anderson, J. R. Ensher, M. R. Matthews,
C. E. Wieman, E. A. Cornell, \textit{Observation of Bose-Einstein
condensation in a atomic vapor}, Science \textbf{269}, 198 (1995);
C. C. Bradley, C. A. Sackett, J. J. Tollett, R. G. Hulet,
\textit{Evidence of Bose-Einstein condensation in a atomic gas
with attractive interactions}, Phys. Rev. Lett. \textbf{75}, 1687
(1995); K. B. Davis, M. O. Mewes, M. R. Andrews, N. J. van Druten,
D. S. Durfee, D. M. Kurn, W. Ketterle, \textit{Bose-Einstein
condensation in a gas of sodium atoms}, Phys. Rev. Lett.
\textbf{75}, 3969 (1995).
\bibitem{wei} S. Weinberg, \textit{Gravitation and cosmology: principles and applications of the
General Theory of Relativity}, Wiley, New York, 1972.
\bibitem{Bar} C. Barcel\'{o}, S. Liberati, M. Visser, \textit{Analogue model for
FRW cosmologies}, Int. J. Mod. Phys. D \textbf{12}, 1641 (2003);
C. Barcel\'{o}, S. Liberati, M. Visser, \textit{Probing
semiclassical analogue gravity in Bose-Einstein condensates with
widely tunable interactions}, Phys. Rev. A \textbf{68}, 053613
(2003).
\bibitem{lan} L. D. Landau, E. M. Lifshitz, \textit{The classical theory of fields}, Butterworth-Heinemann, Amsterdam, 1994.
\bibitem{landau} L. D. Landau, \textit{Theory of superfluidity of helium-II}, J. Phys. USSR \textbf{5}, 71 (1941).
\bibitem{stone} M. Stone, \textit{Acustic energy and momentum in a moving
medium}, Phys. Rev. E \textbf{62}, 1341 (2000).
\bibitem{noi} A. Naddeo, G. Scelza, work in preparation.
\end{thebibliography}
\end{document}